\documentclass[twocolumn,amsmath,amssymb]{revtex4-1}

\usepackage{graphicx}
\usepackage{dcolumn}
\usepackage{bm}
\usepackage[switch,columnwise]{lineno}
\newcommand{\etal}{\emph{et al.}}
\newcommand{\be}{\begin{equation}}
\newcommand{\ee}{\end{equation}}
\newcommand{\bfig}{\begin{figure}}
\newcommand{\efig}{\end{figure}}

\usepackage{lineno}
\usepackage{lipsum}

\begin{document}

\title{Phonon thermal transport shaped by strong spin-phonon scattering in a Kitaev material Na$_2$Co$_2$TeO$_6$}

\author{Xiaochen Hong$^{1,2,*}$}
\author{Matthias Gillig$^{2}$}
\author{Weiliang Yao$^{3}$}
\author{Lukas Janssen$^{4,\dagger}$}
\author{Vilmos Kocsis$^{2}$}
\author{Sebastian Gass$^{2}$}
\author{Yuan Li$^{3,5}$}
\author{Anja U. B. Wolter$^{2}$}
\author{Bernd B\"{u}chner$^{2,6}$}
\author{Christian Hess$^{1,2,\ddagger}$}

\affiliation{
$^1$Fakult\"{a}t f\"{u}r Mathematik und Naturwissenschaften, Bergische Universit\"{a}t Wuppertal, 42097 Wuppertal, Germany\\
$^2$Leibniz-Institute for Solid State and Materials Research (IFW-Dresden), 01069 Dresden, Germany\\
$^3$International Center for Quantum Materials, School of Physics, Peking University, 100871 Beijing, China\\
$^4$Institut f\"{u}r Theoretische Physik and W\"{u}rzburg-Dresden Cluster of Excellence $ct.qmat$, Technische Universit\"{a}t Dresden, 01062 Dresden, Germany\\
$^5$Collaborative Innovation Center of Quantum Matter, 100871 Beijing, China\\
$^6$Institute of Solid State and Materials Physics and W\"{u}rzburg-Dresden Cluster of Excellence $ct.qmat$, Technische Universit\"{a}t Dresden, 01062 Dresden, Germany
}

\maketitle

{\bf
The recent report of a half-quantized thermal Hall effect in the Kitaev material $\alpha$-RuCl$_3$ has sparked a strong debate on whether it is generated by Majorana fermion edge currents or whether other more conventional mechanisms involving magnons or phonons are at its origin. A more direct evidence for Majorana fermions which could be expected to arise from a contribution to the longitudinal heat conductivity $\kappa_{xx}$ at $T\rightarrow0$ is elusive due to a very complex magnetic field dependence of $\kappa_{xx}$.
Here, we report very low temperature (below 1~K) thermal conductivity ($\kappa$) of another candidate Kitaev material, Na$_2$Co$_2$TeO$_6$. The application of a magnetic field along different principal axes of the crystal reveals a strong directional-dependent magnetic-field ($\bf B$) impact on $\kappa$.
We show that no evidence for mobile quasiparticles except phonons can be concluded at any field from 0~T to the field polarized state.
In particular, severely scattered phonon transport is observed across the $B-T$ phase diagram, which is attributed to prominent magnetic fluctuations.
Cascades of phase transitions are uncovered for all $\bf B$ directions by probing the strength of magnetic fluctuations via a precise record of $\kappa$($B$).
Our results thus rule out recent proposals for itinerant magnetic excitations in Na$_2$Co$_2$TeO$_6$, and emphasise the importance of discriminating true spin liquid transport properties from scattered phonons in candidate materials.
}

\vspace{3mm}\noindent

\begin{figure*}
\includegraphics[clip,width=0.99\textwidth]{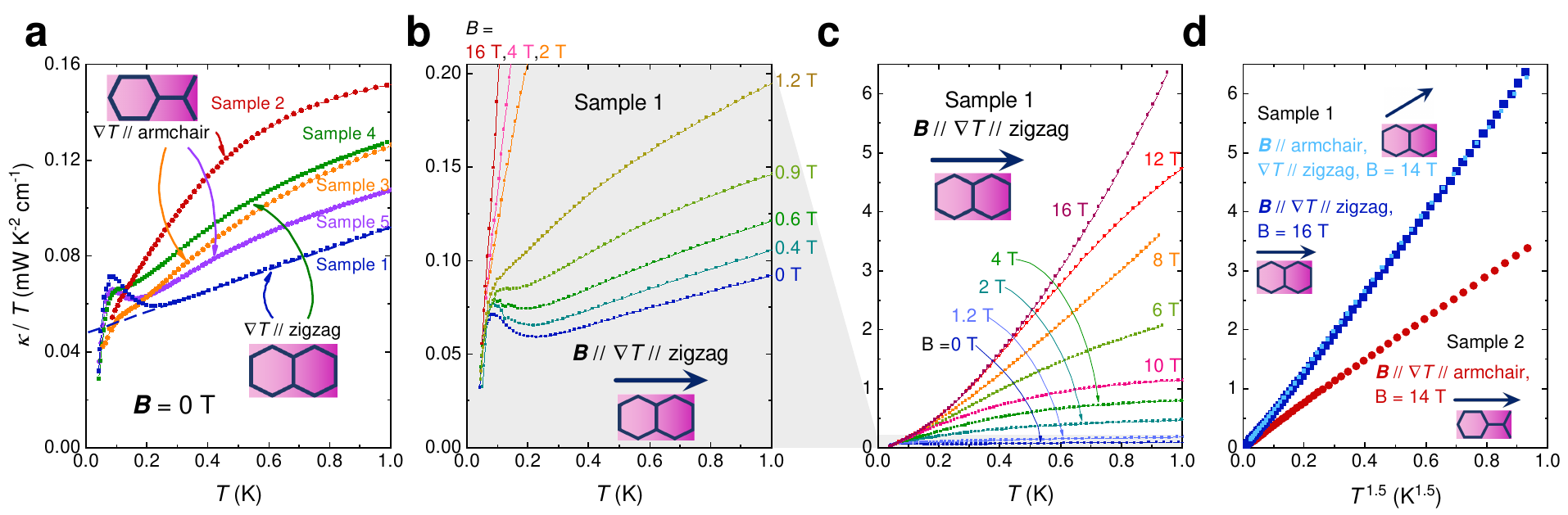}
\caption{{\bf Thermal conductivity of Na$_2$Co$_2$TeO$_6$ and the impact of the in-plane magnetic fields.}
{\bf a,} In-plane thermal conductivity of five Na$_2$Co$_2$TeO$_6$ crystals without magnetic field. The thermal gradient ($\nabla T$) was set $\bf \parallel$ $zigzag$ for Sample\#1 and Sample\#4, and $\bf \parallel$ $armchair$ for the other three samples. By extrapolating the data to $T =$ 0~K, a residual linear term ($\kappa/T|_{T\rightarrow0}$) can be resolved for all five $\kappa/T$($T$) curves, especially clear for Sample\#1, as indicated by the dashed blue line. 
{\bf b,} The field dependence of the $\kappa/T$($T$) curves for Sample\#1, focusing on the small field limit which covers a small portion of {\bf c,} the full field dependence of the $\kappa/T$($T$) curves, as indicated by the gray background. At $T =$~1 K, the $\kappa$ of Sample\#1 experience a 70-fold increase from $B =$~0 T to the saturation field.
{\bf d,} The $\kappa/T$($T$) curves display a typical phononic $\propto T^\alpha$ behavior with $\alpha =$~1.5 at high enough fields. The field direction does not affect the saturated $\kappa/T$($T$) curves.
}
\end{figure*}

Frustrated magnets refer to a set of materials that do not order down to temperatures well below the corresponding magnetic interaction strength.
In some cases, the system is left with a ground state called quantum spin liquid (QSL), which is characterized by entangled spins, topological orders, and fractionalised magnetic excitations \cite{Balents,Savary,Zhou}.
Kitaev QSL, proposed for bond-dependent nearest-neighbour interacting spins on the two-dimensional honeycomb lattice \cite{Kitaev}, is among the rare examples of exactly solvable QSL models \cite{YaoH,McGreevy,Wen,Moessner}. 
Recent research interest is further fueled by possible achieving Kitaev QSL in realistic materials. In particular, fingerprints of novel QSL-related excitations were revealed in $\alpha$-RuCl$_3$ \cite{Takagi,Broholm}, both in its zero-field phase above the magnetic order \cite{Banerjee}, and more intriguingly, in its field-induced quantum disordered state \cite{Kasahara}.

However, those sought-after signatures for QSLs observed in $\alpha$-RuCl$_3$ always come along with alternative explanations.
To be more specific on transport experiments, the report of half-integer quantization of thermal Hall conductance was once regarded as a definitive proof for the existence of Majorana fermions in the field-induced quantum disordered state in  $\alpha$-RuCl$_3$ \cite{Kasahara}.
However, following experiments noticed such quantisation is less reproducible, and attributed the non-quantised temperature-dependent thermal Hall effect to phonons \cite{Lefrancois}, or topological magnons \cite{CzajkaNM}.
On the other hand, quantum oscillations of longitudinal thermal conductivity in the quantum disordered state of $\alpha$-RuCl$_3$ suggest some fractionalised magnetic excitations forming a Fermi surface \cite{CzajkaNP}.
But such spectacle is also argued to be a result simply due to scattered phonons \cite{Lefrancois}, in particular taking into account that no itinerant magnetic excitation has been found in a previous low temperature work \cite{YuYJ}. 
On top of all these inconsistencies are the facts that $\alpha$-RuCl$_3$ suffers from the notorious stacking fault problem \cite{ZhangHD}, from the dechlorination and oxidation after heat treatment \cite{Breitner}, and more generally from sample dependence \cite{Kasa}.
Thus, it is indispensable to seek for another example that can contribute to clarify to what extent these results are intrinsic to a Kitaev material.

In this paper, we studied the ground state properties of a quantum magnet Na$_2$Co$_2$TeO$_6$ (NCTO), which was recently investigated as a candidate Kitaev QSL \cite{LiuHM, Lin, Hong}, through the prism of low temperature thermal conductivity. 
Similar to its more famous cousin $\alpha$-RuCl$_3$, NCTO also features an intermediate quantum disordered phase sandwiched between the low field magnetically ordered state and the high-field polarized state \cite{Lin, Hong}, where a QSL state is prospected.
In this work, we report an anisotropic and astonishing large $\bf B$ impact on $\kappa$ at low temperatures ($< 1$~K). The analysis of our data demonstrates that there is $no$ evidence for mobile gapless QSL excitations ($\kappa_{QSL}$) contributing to $\kappa$. Instead, strongly scattered phonon transport ($\kappa_{ph}$), a quite traditional mechanism, is responsible for all the observed phenomena.
Utilizing the high resolution $\kappa(B)$ data, we reveal a large density of magnetic excitiations in the zero-field ground state of NCTO, further promoted excitations by magnetic field perpendicular to the honeycomb plane, and possible quantum criticality driven by in-plane magnetic field. 

The residual linear term of thermal conductivity extrapolated to the low temperature limit ($\kappa_0/T \equiv \kappa/T|_{T\rightarrow0}$) is persistently scrutinized for QSL candidates \cite{Zhou,Yamashita2012}, since its existence is practically regarded as a criterion for itinerant fractionalised magnetic excitations \cite{Li2020,PanBY,Hartstein,Yamashita2010,Sato}.
Fig.~1a shows $\kappa/T$ of five NCTO crystals in zero magnetic field as a function of $T$.
Albeit certain sample dependence, the $\kappa/T(T)$ profiles of all investigated crystals do not follow the simple power-law behavior expected for common insulators.
Instead, regardless of the direction along which the thermal gradient $\nabla T$ is established (the two in-plane principal axes for the Co-honeycomb lattice, denoted as $zigzag$ and $armchair$ directions, indicated by the schematic insets), all $\kappa/T(T)$ profiles trend to end with a finite $\kappa_0/T$.

Among them, Sample\#1 is especially notable for displaying a pronounced hump which emerges below about 250~mK.
$\kappa_0/T$ of Sample\#1, extracted from a linear fit to its data above 250~mK, as highlighted by the dashed line in Fig.~1a, is 0.048 mW/K$^{-2}$cm$^{-1}$.
At first glance, this finite $\kappa_0/T$ is consistent with the notion that the ground state of NCTO is a QSL, since $\kappa_{QSL}$ contributed by the putative fractionalized magnetic excitations is expected to scale linearly with $T$ at the lowest temperatures \cite{Yamashita2012}. 
We are aware of a recent work reporting similar data of NCTO, where these results have been interpreted as evidences for $\kappa_{QSL}$ \cite{Guang}. 
However, as clearly shown in Fig.~1a, the upturn of $\kappa/T(T)$ for Sample\#1 (same for Sample\#4 and Sample\#5) terminates abruptly below about 100 mK, followed by a sharp drop, which results in a vanishing residual linear term that is incompatible with a gapless QSL ground state.
Furthermore, we argue such observation is at odds with a hypothesised gapped QSL state with a gap size of the order $\sim$ 100~mK. 
When a small in-plane magnetic field is applied, $\kappa_{QSL}/T$($T$) at temperatures below the initial gap energy scale is expected to increase due to a proliferation of quasiparticle excitations, while $\kappa/T$($T$) at higher temperatures should be less sensitive to the field since $\kappa_{ph}$ is not directly affected.
This is just opposite to our observations. As depicted in Fig.~1b, the field changes $\kappa/T$($T$) fundamentally above 100~mK, while it has basically no impact on the data at the lowest temperature.

\begin{figure*}
\includegraphics[clip,width=0.99\textwidth]{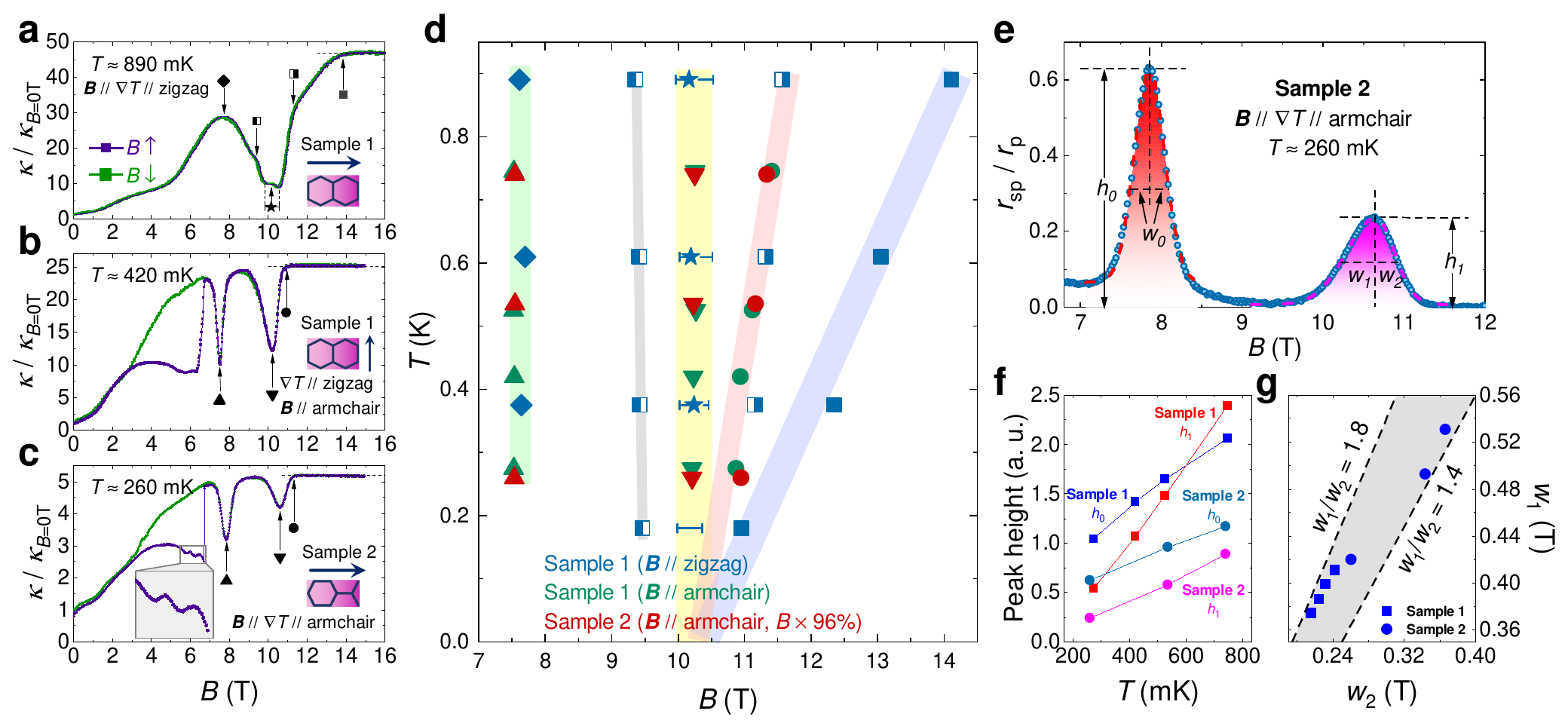}
\caption{{\bf Anisotropic in-plane magnetic field effects on the thermal conductivity.}
Panels {\bf a, b,} and {\bf c,} plot the normalized $\kappa$($B$) profiles of different combinations of ($\nabla T$, $B$) orientations at selected temperatures. 
The purple curves and the green curves depict the data collected with field increasing and decreasing, respectively. All $\kappa$($B$) curves experience a saturation at high fields, signaling that a polarized state was reached in this study.
Panels {\bf a} and {\bf b} depict the results of the same crystal (Sample\#1) with the same $\nabla T$ orientation ($\nabla T\parallel zigzag$) but with different $\bf B$ directions. They behave fundamentally different.
Meanwhile Panel {\bf b} resembles Panel {\bf c} which is the data collected from another crystal (Sample\#2) with a different $\nabla T$ orientation ($\nabla T\parallel armchair$) but the same field direction ($\bf B\parallel$ $armchair$). As highlighted in the inset of Panel {\bf c}, an oscillation-like feature can be distinguished before an abrupt increase of the $\kappa$($B$) for the field-up ramp.
The anomalies in all $\kappa/T$($B$) curves are marked by different symbols.
{\bf d,} The temperature evolution of these anomalies of $\kappa$($B$) profiles. They are grouped into different categories as glided by the background stripes. The field values for Sample\#2 are multiplied by 96\%, in order to compromise the possible (out-of-plane) misalignment.
{\bf e,} A representative case of analysing the renormalized phonon-spin scattering strength ($r_{sp}$/$r_p$, see text). The peak at 7.5~T and 10.2~T (the field values for Sample\#2 are multiplied by 96\% as mentioned above) are fit to standard Gaussian function and skewed Gaussian function, respectively. 
The height and span of the peaks ($h_0$, $w_0$ for the 7.5~T peak, $h_1$, $w_1$, and $w_2$ for the 10.2~T peak) are extracted as shown in the figure, and are summarized in Panel {\bf f} and Panel {\bf g}, respectively.
}
\end{figure*}

Higher magnetic fields have a drastic impact on the $\kappa(T)/T$ curves until they finally saturate into a power-law behavior over the whole temperature range at the highest field.
A representative case of Sample\#1 is shown in Fig.~1c, along with the others displayed in Supp. Note 1 \cite{SM}. 
As can be seen in Fig.~1d, regardless of the (in-plane) field directions, $\kappa(T)/T$ curves for a given sample saturate at the same value.

Recent high-field magnetization experiments found the field-polarised state is reached for in-plane magnetic field higher than 10~T in NCTO \cite{Zhang}.
As a result, spins can not contribute anymore, neither directly, nor indirectly through scattering at the highest field reached in this study.
The saturated $\kappa(T)/T$ thus represents a standard ballistic phonon heat conductivity $\kappa_{ph}$ in the low-temperature limit with $\kappa_{ph}/T \propto T^\alpha$, of which the exponent $\alpha$ = 1.5.
The sample independent $\alpha < 2$ indicates that the phonon reflection at the sample surface is not perfectly diffuse, and should be attributed to the intrinsic surface roughness of the as-grown NCTO crystals.
One can thus take the $\kappa(T)/T$ curves at highest in-plane field as the intrinsic phonon thermal conductivity $\kappa_{ph}/T(T)$ free from the spin-phonon scattering.

As can be concluded from Fig.~1 and Supp. Fig.~1, there is no evidence for finite $\kappa_0/T$ at any field. Furthermore, $\kappa/T(T)$ profiles at zero and low fields never exceed the saturated $\kappa_{ph}/T(T)$ over the whole temperature range.
Hence, any evidence for a direct magnetic thermal transport channel on top of $\kappa_{ph}$, including $\kappa_{QSL}$, is elusive. If present, it must be negligibly small since the total measured $\kappa$ at zero field is already at a very small value and at finite fields the saturated $\kappa_{ph}$ is never exceeded.
Hence, the intricate landscape of $\kappa(T)/T$ and its field dependence can be reasonably accounted for by strong phonon-spin scattering, which is the main theme of the rest of the paper.

The field dependence of $\kappa/T$($T$) is non-monotonic, see Fig.~1c.
A closer inspection can be performed by isothermal $\kappa/T$($B$) measurements.
Fig.~2a-c exhibit the representative data of two samples with three different ($\nabla T$, $\bf B$) direction combinations. 
A straightforward conclusion is that $\kappa(B)$ is dictated by the $\bf B$ direction, irrelevant to the $\nabla T$ direction. That further underlines the dominance of scattered $\kappa_{ph}$ in NCTO over the whole field-temperature phase space.
More specifically, a magnetic field applied along the $zigzag$ and $armchair$ directions on the same sample results in sharply different $\kappa$($B$) profiles.
For $\bf B\parallel$ $zigzag$ (Fig.~2a), there is strictly $no$ hysteresis behavior. The overall feature turns out to be a broad dip centered around 10.2~T, with many minor anomalies marked by different symbols in the figure.
For $\bf B\parallel$ $armchair$ (Fig.~2b, Fig.~2c and Supp. Fig.~2e-2k), there are two very sharp dips at 7.5~T and 10.2~T.
In contrast to the $\bf B\parallel$ $zigzag$ case, a large hysteresis between 3~T and 6.5~T is evident. The abrupt increase of $\kappa$ in the up-ramp isotherm points to a first-order-like transition.
Note that the $\bf B$ direction dependent occurrence of hysteresis was already mentioned by higher temperature investigations \cite{Hong,Yao2020}.
More discussion on the hysteretic behavior can be found in the Supp. Note 2 \cite{SM}.
With both $\bf B$ directions, $\kappa$ saturates at high enough fields, as indicated by the dashed lines.

\begin{figure*}
\includegraphics[clip,width=0.9\textwidth]{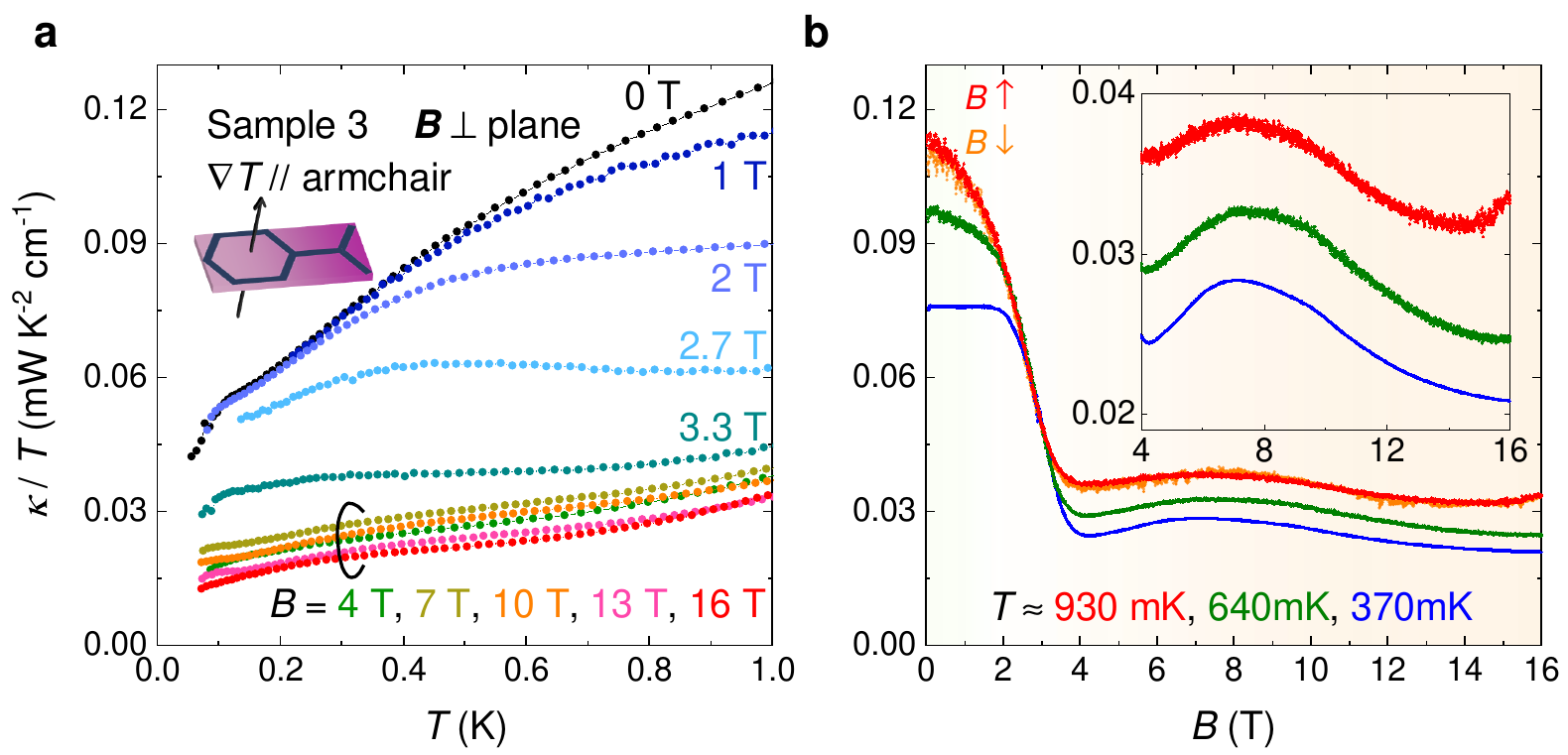}
\caption{ 
{\bf Further suppression of $\kappa$ by out-of-plane magnetic fields.}
{\bf a,} The in-plane $\kappa$ of Sample\#3 under out-of-plane magnetic fields. 
The geometry is sketched in the inset. $\nabla T$ is along the $armchair$ direction, and $\bf B$ is perpendicular to the honeycomb plane.
A rapid field suppression of $\kappa/T$ is recognized at finite temperatures below about 3~T. At higher fields, the $\kappa/T$($T$) curves changes nonmonotonically in a narrow range.
{\bf b,} The $\kappa/T$($B$) profiles at three representative temperatures. They clearly show a fast drop at field 2~T $< B <$ 4~T, followed by an oscillation-like feature at higher fields. As highlighted in the inset, the oscillation profiles change little with temperature.}
\end{figure*}

All the salient features in Fig.~2a-c and Supp. Fig.~2 are summarized in Fig.~2d.
Beside the saturation field, which increases roughy linearly with temperature, the other anomalies are basically temperature independent.
Such rich phase diagram suggests the landscape of successive $\bf B$ induced phase transitions of the Heisenberg-Kitaev model might be inherited by NCTO \cite{Janssen}. 
Remarkably, three different finite-temperature anomalies marked by the yellow, red, and blue shaded areas seem to merge at around 10.2~T at 0~K, suggesting a quantum critical point.
We expect the presented result to motivate further theoretical studies that account for specified parameters of NCTO in order to demystify these phases. 

Some theoretical exploration on the Heisenberg-Kitaev honeycomb model have already revealed the potentially very rich phase diagram tuned by magnetic field \cite{Janssen}.
Of particular interest is a multi-Q phase that can be stabilized in finite fields \cite{Janssen}.
Recent experiments proposed the zero-field ground state of NCTO could be a triple-Q order \cite{Chen,Lee,Kruger,Yao2022,Kikuchi}, potentially stabilized by the proposed ring exchange interactions \cite{Kruger}. 
If confirmed, it could be a natural explanation of the extensive scattering of $\kappa_{ph}$ at zero field, since a triple-Q ordered state guarantees larger density of state (DOS) of excitations around zero momentum ($\Gamma$ point) which interact strongly with phonons \cite{Kruger}.
These magnetic excitations should in principle contribute to specific heat. A decent experiment to exam this property is still pending for NCTO since heat capacity measurement at very low temperature is technically challenging \cite{Wilhelm}.
Nevertheless, in Supp. Note 3 we argue our raw data can provide some indirect evidences for huge specific heat in the zero-field ground state \cite{SM}.
Our data thus corroborate the notion of very large magnetic DOS as is expected for a triple-Q state.

Next we comment on the oscillation-like $\kappa$($B$) in the intermediate field range when $\bf B\parallel$ $armchair$, as highlighted in the inset of Fig.~2c.
A similar feature exists but is less obvious in another sample and at higher temperatures, see Supp. Fig.~2g, 2h, and 2j \cite{SM}.
Gapless QSL states that foster quantum oscillation of presumably charge neutral magnetic excitations are indeed anticipated for the Kitaev model \cite{Kitaev,Gohlke,Hickey}, and was invoked to interpret the $\kappa$($B$) oscillations in $\alpha$-RuCl$_3$ \cite{CzajkaNP}.
Nevertheless, we believe quantum oscillations of $\kappa_{QSL}$ should be irrelevant to NCTO.
First of all, the oscillation feature in NCTO exists in the hysteresis region, thus in a magnetically ordered state.
Besides, the fact that there is no evidence for itinerant $\kappa_{QSL}$ at any field does not favor a quantum oscillation picture. 
Additionally, it is hard to imagine a significant direct contribution of $\kappa_{QSL}$ which is exempted from the strong fluctuations that results in a reduction of $\kappa_{ph}$ by orders of magnitude.
With this in mind, more caution should be taken for uniqueness of possible Kitaev QSL signatures observed in this material class, since thermal transport properties of NCTO share impressively similar features with $\alpha$-RuCl$_3$ \cite{Hentrich,Hentrich2,Hong,Gillig}.

In view of the scattered phononic thermal conductivity has been revealed for NCTO, it is interesting to utilize the field evolution of $\kappa/T$ as a natural probe for the phonon-spin scattering in this material \cite{Linarite, YCOB}.
The relatively simple profiles of the $\bf B\parallel$ $armchair$ cases (Fig.~2b and~2c) allow a straightforward analysis.
According to Matthiessen's rule, the total phonon scattering rate is the sum of all independent ingredients, in this case the phonon-spin scattering ($r_{sp}$) and the intrinsic phononic scattering ($r_p$).
At very low temperatures, $r_p$ is determined only by the crystal size and is temperature independent because the phonon wave length diverges. It is the only scattering mechanism at play in the field polarized phase in the temperature range covered by this study.
So the renormalized $r_{sp}$/$r_p$ ($\equiv \kappa_{ph}/\kappa(B)-$1) is an index reflecting the magnetic scattering strength \cite{YCOB}.
Fig.~2e exhibits one representative case for Sample\#2 at about 260~mK. The $B = 7.5$~T peak is perfectly reproduced by a standard Gaussian distribution, which indicates randomly distributed disorder.
However, the $B = 10.2$~T peak can only be reasonably fit by a skewed Gaussian distribution with different width at half maximum on both sides ($w_1 \neq w_2$).
The extracted parameters are summarized in Fig.~2f and Fig.~2g.
The fact that $w_1 > w_2$ might be key for identifying the actual type of phases surrounding the 10.2~T criticality, which needs material specific calculations to unravel.

Finally, at low temperatures, when $\bf B$ is applied perpendicular to the NCTO honeycomb plane, $\kappa$ gets further suppressed beyond a critical field of $B \approx 3$~T, as shown in Fig.~3a. 
In the language of magnetic scattering, it means the already strong magnetic fluctuations at zero field are even further promoted, contradictory to the intuition that field alignment of the magnetic moments should reduce them.

Indeed, there are some theoretical considerations predicting $\bf B\parallel$ $c$ can lead to abundant phase transitions in the original Kitaev model and its variants \cite{Li,Sorensen,Chern,Gohlke,Janssen}. 
Especially, quantum fluctuations in some of these intermediate phases are expected to be rather strong \cite{Chern}, in line with our observations that $\kappa_{ph}$ gets further suppressed compared to the zero-field phase. 
As displayed more clearly in Fig.~3b, $\kappa/T$($B$) experiences a rapid drop between 2~T and 4~T.
It is non-monotonic at higher field, featured by a broad hump around 8~T and a broad dip around 15~T, highlighted by the inset.
We note the anomaly at about $15$~T can be related to phase transitions revealed by a recent magnetization work \cite{Zhang}, while the rapid drop between $2$~T and $4$~T do not have a similar correspondence.

To conclude, our high-quality low-temperature thermal transport study on NCTO single crystals established the dominance of phonon thermal transport in its magnetic ground states, and exclude itinerant magnetic excitations of any form. The field evolution of $\kappa$, which manifests the strength of phonon-spin scattering, was taken as a probe to elucidate the intricate magnetic phases in NCTO.
The strongly scattered phonons in the zero-field ground state indicates large DOS of magnetic excitations, in line with the proposal of a triple-Q state in NCTO.
The in-plane $\kappa$ depends significantly of the in-plane field direction, underlines the bond-dependent interactions of a Kitaev material.
However, a criticality at $10.2$~T is present regardless of the in-plane field direction.
Field perpendicular to the honeycomb plane also induces multiple magnetic phases, where phonon-spin scattering is even more prominent than the zero-field phase.
We expect similar physics should be quite common among QSL candidate materials, since QSLs are intrinsically at the edge of competing orders. The pervasive phase boundaries should be careful considered when analysing existing and future experimental results of QSL candidate materials.

\vspace{3mm}
\noindent
{\bf Methods}\\
{\bf Sample preparation}\\
Single crystals of Na$_2$Co$_2$TeO$_6$ were prepared with a modified flux method. Na$_2$CO$_3$, Co$_3$O$_4$ and TeO$_2$ powders were grounded and loaded into an alumina crucible. Excess amount of TeO$_2$ served as a self-flux. The crucible was heated up to 1050$^{\circ}$C, kept for two days, before cooling down to 600$^{\circ}$C at a rate of 6.5$^{\circ}$C per hour.
Ruby-colored hexagonal flakes of typical size $\sim$10$\times$10$\times$0.1 mm$^3$ can be mechanically collected out of
bluish violet residue. The harvested single crystals were further washed with a NaOH solution.
Basic characterizations of the as-grown crystals can be found in Ref~\cite{Yao2020}.\\
The samples were cut into a size of about $\sim$5$\times$1$\times$0.1 mm$^3$, the longest direction along which $\nabla T$ was generated are the $armchair$ or $zigzag$ directions, respectively.
In order to make thermal contacts to the fresh surface, the out-most layers of the as-grown crystals were cleaved off by a blade just before the silver paint was glued to it. As a result, the thickness of the samples were reduced to about 60 $\mu$m.\\
{\bf Heat transport measurements}\\
The thermal conductivities were measured in a dilution refrigerator, using a standard four-wire steady-state method with two RuO$_2$ chip thermometers, calibrated $in$ $situ$ against a reference RuO$_2$ thermometer.
For fixed field $\kappa(T)$ measurements, each sample was initially cooled down from room temperature without magnetic field. The data were collected with gradually increased magnetic field in order to avoid the complicated situation of involving the hysteresis effect. $\kappa(T)$ data were collected in a steady-state manner (see Supp. Fig.~3 \cite{SM}).\\
To measure the $\kappa(B)$ isotherms, the system was cooled from the paramagnetic phase in zero field. Then the sample temperature was kept at a set point, and the field was changed from 0~T to the highest field (up-ramp), followed by decreasing to 0~T (down-ramp). The field was changed at a speed no more than 20~mT per minute, in order to minimize its heating effect, and to keep the system in a (quasi-)thermal equilibrium state.
In some cases, the field were ramped up and down multiple times, see Supp. Note 2 for more details \cite{SM}.\\   
We wish to point out that according to our experience, NCTO seems to have very strong anisotropic in-plane magnetic susceptibility at very low temperatures, that tend to align the $armchair$ direction to the field. 
We are aware of a recently posted paper which reports a direct measurement of this magnetic anisotropy that underpins our conjecture \cite{Lin2022}.
In order to avoid experimental artifacts caused by such effect, the cooler side of all samples were glued to the heat sink (a gilded silver bulk) directly with a strong epoxy (Wakefield DeltaBond-152 two component adhesives). The samples and the wires attached to them were checked carefully under a microscope after the experiments, to confirm they had not bent during the measurements.\\

{\bf Thermodynamic measurements}\\
The magnetization measurements were carried out using a SQUID magnetometer (iHelium3, MPMS-XL, Quantum Design). The magnetostriction was measured using a commercially available dilatometer (Standard probe, Kuechler) compatible with the Quantum Design PPMS systems. These measurements are based on the capacitance measurement technique (AH2700A, Andeen-Hagerling).

\vspace{1cm}
\centerline{* ~~~ * ~~~  *}

\vspace{1cm}\noindent
Corresponding author email: \\
$^{*}$xhong@uni-wuppertal.de\\
$^{\dagger}$lukas.janssen@tu-dresden.de\\
$^{\ddagger}$c.hess@uni-wuppertal.de

\vspace{5mm}\noindent
{\bf Acknowledgements} \\
We thank Jeroen van den Brink, Matthias Vojta, Vladislav Kataev, Christoph Wellm, Satoshi Nishimoto, Wenjie Chen, Yang Xu, and Wilhelm Kr\"{u}ger for illuminating discussions and collaboration on related work.
We would like to thank Nicol\'{a}s P\'{e}rez for generously sharing the equipment, and thank Danny Baumann and Tino Schreiner for their technical assistance, especially during the pandemic lab-shutdown crisis.\\
{\bf Funding Acknowledgement:}\\
This work has been supported by the Deutsche Forschungsgemeinschaft (DFG) through SFB 1143 (Project-ID No.247310070) and the W\"{u}rzburg-Dresden Cluster of Excellence $ct.qmat$ (EXC 2147, Project-ID No. 390858490).\\
C.H. has received funding from the European Research Council (ERC) under the European Union's-Horizon 2020 research and innovation programme (Grant Agreement No. 647276-MARS-ERC-2014-CoG).\\
L.J. was supported by the DFG through the Emmy Noether program (JA2306/4-1, Project No. 411750675).\\
V.K. was supported by the Alexander von Humboldt Foundation.\\
Y.L. acknowledges funding support from the National Basic Research Program of China (Grant No. 2021YFA1401901) and the NSF of China (Grant No. 12061131004).

\vspace{3mm}
\noindent
{\bf Author contributions}\\
B.B., C.H., and Y.L. conceived and initiated the study. W.Y. and Y.L. grown the crystals. X.H. and M.G. performed the transport
measurements. V.K., S.G., and A.U.B.W. performed the thermodynamic measurements.
X.H., C.H., and L.J. analysed the data.  X.H., C.H., L.J., and V.K. wrote the manuscript with input from all authors. C.H.
supervised the project.

\vspace{3mm}
\noindent
{\bf Additional Information}\\
Supplementary information is available at the end of the paper.

\vspace{3mm}
\noindent
{\bf Competing financial interests}\\
The authors declare no competing financial interests.

\newcommand{\addpage}[1] {
 \begin{figure*}
   \includegraphics[width=7in,page=#1]{NCTO_LT_SM.pdf}
 \end{figure*}
}
\addpage{1}
\addpage{2}
\addpage{3}
\addpage{4}
\addpage{5}
\addpage{6}


\begin{thebibliography}{0}%
\makeatletter
\providecommand \@ifxundefined [1]{%
 \@ifx{#1\undefined}
}%
\providecommand \@ifnum [1]{%
 \ifnum #1\expandafter \@firstoftwo
 \else \expandafter \@secondoftwo
 \fi
}%
\providecommand \@ifx [1]{%
 \ifx #1\expandafter \@firstoftwo
 \else \expandafter \@secondoftwo
 \fi
}%
\providecommand \natexlab [1]{#1}%
\providecommand \enquote  [1]{``#1''}%
\providecommand \bibnamefont  [1]{#1}%
\providecommand \bibfnamefont [1]{#1}%
\providecommand \citenamefont [1]{#1}%
\providecommand \href@noop [0]{\@secondoftwo}%
\providecommand \href [0]{\begingroup \@sanitize@url \@href}%
\providecommand \@href[1]{\@@startlink{#1}\@@href}%
\providecommand \@@href[1]{\endgroup#1\@@endlink}%
\providecommand \@sanitize@url [0]{\catcode `\\12\catcode `\$12\catcode
  `\&12\catcode `\#12\catcode `\^12\catcode `\_12\catcode `\%12\relax}%
\providecommand \@@startlink[1]{}%
\providecommand \@@endlink[0]{}%
\providecommand \url  [0]{\begingroup\@sanitize@url \@url }%
\providecommand \@url [1]{\endgroup\@href {#1}{\urlprefix }}%
\providecommand \urlprefix  [0]{URL }%
\providecommand \Eprint [0]{\href }%
\providecommand \doibase [0]{http://dx.doi.org/}%
\providecommand \selectlanguage [0]{\@gobble}%
\providecommand \bibinfo  [0]{\@secondoftwo}%
\providecommand \bibfield  [0]{\@secondoftwo}%
\providecommand \translation [1]{[#1]}%
\providecommand \BibitemOpen [0]{}%
\providecommand \bibitemStop [0]{}%
\providecommand \bibitemNoStop [0]{.\EOS\space}%
\providecommand \EOS [0]{\spacefactor3000\relax}%
\providecommand \BibitemShut  [1]{\csname bibitem#1\endcsname}%
\let\auto@bib@innerbib\@empty
\end{thebibliography}%


\begin{thebibliography}{88}

\bibitem{Balents} Balents, L., 
Spin liquids in frustrated magnets.
\emph{Nature} {\bf 464}, 199–208 (2010). 
https://doi.org/10.1038/nature08917

\bibitem{Savary} Savary, L. $\&$ Balents, L.,
Quantum spin liquids: a review.
\emph{Rep. Prog. Phys.} {\bf 80}, 106502 (2017).
DOI 10.1088/0034-4885/80/1/016502

\bibitem{Zhou} Zhou, Y., Kanoda, K. $\&$ Ng, T.K.
Quantum spin liquid states.
\emph{Rev. Mod. Phys.} {\bf 89}, 025003 (2017).
https://doi.org/10.1103/RevModPhys.89.025003

\bibitem{Kitaev} Kitaev, A.,
Anyons in an exactly solved model and beyond.
\emph{Ann. Phys.} {\bf 321}, 2-111 (2006).
https://doi.org/10.1016/j.aop.2005.10.005

\bibitem{Moessner} Moessner, R., $\&$ Sondhi, S. L.,
Resonating Valence Bond Phase in the Triangular Lattice Quantum Dimer Model.
\emph{Phys. Rev. Lett.} {\bf 86}, 1881 (2001).
https://doi.org/10.1103/PhysRevLett.86.1881

\bibitem{Wen} Wen, X.-G.,
Quantum Orders in an Exact Soluble Model.
\emph{Phys. Rev. Lett.} {\bf 90}, 016803 (2003).
https://doi.org/10.1103/PhysRevLett.90.016803

\bibitem{YaoH} Yao, H., Zhang, S.-C., $\&$ Kivelson, S. A.,
Algebraic Spin Liquid in an Exactly Solvable Spin Model.
\emph{Phys. Rev. Lett.} {\bf 102}, 217202 (2009).
https://doi.org/10.1103/PhysRevLett.102.217202

\bibitem{McGreevy} Ben-Zion, D., Das, D., $\&$ McGreevy, J.,
Exactly solvable models of spin liquids with spinons, and of three-dimensional topological paramagnets.
\emph{Phys. Rev. B} {\bf 93}, 155147 (2016).
https://doi.org/10.1103/PhysRevB.93.155147

\bibitem{Takagi} Takagi, H., Takayama, T., Jackeli, G., Khaliullin, G. $\&$ Nagler, S. E.  
Concept and realization of Kitaev quantum spin liquids.
\emph{Nat. Rev. Phys.} {\bf 1}, 264–280 (2019).
https://doi.org/10.1038/s42254-019-0038-2

\bibitem{Broholm} Broholm, C., Cava, R. J., Kivelson, S. A., Nocera, D. G. Norman, M. R. $\&$ Senthil, T.,
Quantum spin liquids.
\emph{Science} {\bf 367}, 263 (2020).
DOI: 10.1126/science.aay0668

\bibitem{Banerjee} Banerjee, A. \etal, 
Proximate Kitaev quantum spin liquid behavior in a honeycomb magnet.
\emph{Nat. Mat.} {\bf 15}, 733–740 (2016).
https://doi.org/10.1038/nmat4604

\bibitem{Kasahara} Kasahara, Y. \etal,
Majorana quantization and half-integer thermal quantum Hall effect in a Kitaev spin liquid.
\emph{Nature} {\bf 559}, 227–231 (2018).
https://doi.org/10.1038/s41586-018-0274-0

\bibitem{Lefrancois} Lefran\c{c}ois, \'{E}., Baglo, J., Barth\'{e}lemy, Q., Kim, S., Kim, Y.-J. $\&$ Taillefer, L.,
Oscillations in the magnetothermal conductivity of $\alpha$-RuCl$_3$: Evidence of transition anomalies.
\emph{Phys. Rev. B} {\bf 107}, 064408 (2023).
https://doi.org/10.1103/PhysRevB.107.064408

\bibitem{CzajkaNM} Czajka, P. \etal,
Planar thermal Hall effect of topological bosons in the Kitaev magnet $\alpha$-RuCl$_3$.
\emph{Nat. Mater.} {\bf 22}, 36–41 (2023).
https://doi.org/10.1038/s41563-022-01397-w

\bibitem{CzajkaNP} Czajka, P. \etal,
Oscillations of the thermal conductivity in the spin-liquid state of $\alpha$-RuCl$_3$.
\emph{Nat. Phys.} {\bf 17}, 915–919 (2021).
https://doi.org/10.1038/s41567-021-01243-x

\bibitem{YuYJ} Yu, Y.J. \etal,
Ultralow-Temperature Thermal Conductivity of the Kitaev Honeycomb Magnet $\alpha$-RuCl$_3$ across the Field-Induced Phase Transition.
\emph{Phys. Rev. Lett.} {\bf 120}, 067202 (2018).
https://doi.org/10.1103/PhysRevLett.120.067202

\bibitem{ZhangHD} Zhang, H. \etal,
Stacking disorder and thermal transport properties of $\alpha$-RuCl$_3$.
\emph{online preprint} arXiv:2303.03682

\bibitem{Breitner} Breitner, F. A., Jesche, A., Tsurkan, A., $\&$ Gegenwart, P.,
Thermal decomposition of the Kitaev material $\alpha$-RuCl$_3$ and its influence on low-temperature behavior.
\emph{online preprint} arXiv:2303.11308

\bibitem{Kasa} Kasahara, Y. \etal,
Quantized and unquantized thermal Hall conductance of the Kitaev spin liquid candidate $\alpha$-RuCl$_3$.
\emph{Phys. Rev. B} {\bf 106}, L060410 (2022).
https://doi.org/10.1103/PhysRevB.106.L060410

\bibitem{LiuHM} Liu, H. and  $\&$ Khaliullin, G.,
Pseudospin exchange interactions in $d^7$ cobalt compounds: Possible realization of the Kitaev model.
\emph{Phys. Rev. B} {\bf 97}, 014407 (2018).
https://doi.org/10.1103/PhysRevB.97.014407

\bibitem{Lin} Lin, G. \etal,
Field-induced quantum spin disordered state in spin-1/2 honeycomb magnet Na$_2$Co$_2$TeO$_6$.
\emph{Nat. Commun.} {\bf 12}, 5559 (2021).
https://doi.org/10.1038/s41467-021-25567-7

\bibitem{Hong} Hong, X. \etal,
Strongly scattered phonon heat transport of the candidate Kitaev material Na$_2$Co$_2$TeO$_6$.
\emph{Phys. Rev. B} {\bf 104}, 144426 (2021).
https://doi.org/10.1103/PhysRevB.104.144426

\bibitem{Yamashita2012} Yamashita, M., Shibauchi, T. $\&$ Matsuda, Y.,
Thermal-transport studies on two-dimensional quantum spin liquids.
\emph{Chemphyschem} {\bf 13}, 74–78 (2012).
https://doi.org/10.1002/cphc.201100556

\bibitem{Yamashita2010} Yamashita, M. \etal,
Highly Mobile Gapless Excitations in a Two-Dimensional Candidate Quantum Spin Liquid.
\emph{Science} {\bf 328}, 1246-1248 (2010).
DOI: 10.1126/science.118820

\bibitem{Hartstein} Hartstein, M. \etal,
Fermi surface in the absence of a Fermi liquid in the Kondo insulator SmB$_6$.
\emph{Nat. Phys.} {\bf 14}, 166–172 (2018).
https://doi.org/10.1038/nphys4295

\bibitem{Li2020} Li, N. \etal,
Possible itinerant excitations and quantum spin state transitions in the effective spin-1/2 triangular-lattice antiferromagnet Na$_2$BaCo(PO$_4$)$_2$.
\emph{Nat. Commun.} {\bf 11}, 4216 (2020).
https://doi.org/10.1038/s41467-020-18041-3

\bibitem{Sato} Sato, Y. \etal,
Charge-neutral fermions and magnetic field-driven instability in insulating YbIr$_3$Si$_7$.
\emph{Nat. Commun.} {\bf 13}, 394 (2022).
https://doi.org/10.1038/s41467-021-27541-9

\bibitem{PanBY} Pan, B. \etal,
Unambiguous Experimental Verification of Linear-in-Temperature Spinon Thermal Conductivity in an Antiferromagnetic Heisenberg Chain.
\emph{Phys. Rev. Lett.} {\bf 129}, 167201 (2022).
https://doi.org/10.1103/PhysRevLett.129.167201

\bibitem{Guang} Guang, S. K. \etal,
Thermal transport of fractionalized antiferromagnetic and field-induced states in the Kitaev material Na$_2$Co$_2$TeO$_6$.
\emph{Phys. Rev. B} {\bf 107}, 184423 (2023).
https://doi.org/10.1103/PhysRevB.107.184423

\bibitem{SM} Supplementary Materials attached at the end of this file

\bibitem{Zhang} Zhang, S. \etal,
Electronic and magnetic phase diagrams of Kitaev quantum spin liquid candidate Na$_2$Co$_2$TeO$_6$.
\emph{online preprint} arXiv:2212.03849

\bibitem{Yao2020} Yao, W. $\&$ Li, Y.,
Ferrimagnetism and anisotropic phase tunability by magnetic fields in Na$_2$Co$_2$TeO$_6$.
\emph{Phys. Rev. B} {\bf 101}, 085120 (2020).
https://doi.org/10.1103/PhysRevB.101.085120

\bibitem{Janssen} Janssen, L., Andrade, E. C. $\&$ Vojta, M.,
Honeycomb-Lattice Heisenberg-Kitaev Model in a Magnetic Field: Spin Canting, Metamagnetism, and Vortex Crystals.
\emph{Phys. Rev. Lett.} {\bf 117}, 277202 (2016).
https://doi.org/10.1103/PhysRevLett.117.277202

\bibitem{Chen} Chen, W. \etal,
Spin-orbit phase behavior of Na$_2$Co$_2$TeO$_6$ at low temperatures.
\emph{Phys. Rev. B} {\bf 103}, L180404 (2021).
https://doi.org/10.1103/PhysRevB.103.L180404

\bibitem{Lee} Lee, C. H. \etal,
Multistage development of anisotropic magnetic correlations in the Co-based honeycomb lattice Na$_2$Co$_2$TeO$_6$.
\emph{Phys. Rev. B} {\bf 103}, 214447 (2021).
https://doi.org/10.1103/PhysRevB.103.214447

\bibitem{Kruger} Kr\"{u}ger, W. G. F., Chen, W., Jin, X., Li, Y. $\&$ Janssen, L.,
Triple-Q order in Na$_2$Co$_2$TeO$_6$ from proximity to hidden-SU(2)-symmetric point.
\emph{online preprint} arXiv: 2211.16957

\bibitem{Yao2022} Yao, W. \etal,
Magnetic ground state of the Kitaev Na$_2$Co$_2$TeO$_6$ spin liquid candidate.
\emph{Phys. Rev. Research} {\bf 5}, L022045 (2023).
https://doi.org/10.1103/PhysRevResearch.5.L022045

\bibitem{Kikuchi} Kikuchi, J., Kamoda, T., Mera, N., Takahashi, Y., Okumura, K. $\&$ Yasui, Y.,
Field evolution of magnetic phases and spin dynamics in the honeycomb lattice magnet Na$_2$Co$_2$TeO$_6$: $^{23}$Na NMR study.
\emph{Phys. Rev. B} {\bf 106}, 224416 (2022).
https://doi.org/10.1103/PhysRevB.106.224416

\bibitem{Wilhelm} Wilhelm, H., L\"{u}hmann,  T., Rus, T., $\&$  Steglich, F.,
A compensated heat-pulse calorimeter for low temperatures.
\emph{Rev. Sci. Instrum.} {\bf 75}, 2700–270 (2004).
https://doi.org/10.1063/1.1771486

\bibitem{Gohlke} Gohlke, M., Moessner, R. $\&$ Pollmann, F.,
Dynamical and topological properties of the Kitaev model in a [111] magnetic field.
\emph{Phys. Rev. B} {\bf 98}, 014418 (2018).
https://doi.org/10.1103/PhysRevB.98.014418

\bibitem{Hickey} Hickey, C. $\&$ Trebst, S.,
Emergence of a field-driven U(1) spin liquid in the Kitaev honeycomb model.
\emph{Nat. Commun.} {\bf 10}, 530 (2019).
https://doi.org/10.1038/s41467-019-08459-9

\bibitem{Hentrich} Hentrich, R. \etal,
Unusual phonon heat transport in $\alpha$-RuCl$_3$: Strong spin-phonon scattering and field-induced spin gap.
\emph{Phys. Rev. Lett.} {\bf 120}, 117204 (2018).
https://doi.org/10.1103/PhysRevLett.120.117204

\bibitem{Hentrich2} Hentrich, R. \etal,
Large thermal Hall effect in $\alpha$-RuCl$_3$: Evidence for heat transport by Kitaev-Heisenberg paramagnons.
\emph{Phys. Rev. B} {\bf 99}, 085136 (2019).
https://doi.org/10.1103/PhysRevB.99.085136

\bibitem{Gillig} Gillig, M. \etal,
Phononic-magnetic dichotomy of the thermal Hall effect in the Kitaev-Heisenberg candidate material Na$_2$Co$_2$TeO$_6$.
\emph{online preprint} arXiv:2303.03067

\bibitem{Linarite} Gillig, M. \etal,
Thermal transport of the frustrated spin-chain mineral linarite: Magnetic heat transport and strong spin-phonon scattering.
\emph{Phys. Rev. B} {\bf 104}, 235129 (2021).
https://doi.org/10.1103/PhysRevB.104.235129

\bibitem{YCOB} Hong, X. \etal,
Heat transport of the kagome Heisenberg quantum spin liquid candidate YCu$_3$(OH)$_{6.5}$Br$_{2.5}$: Localized magnetic excitations and a putative spin gap.
\emph{Phys. Rev. B} {\bf 106}, L220406 (2022).
https://doi.org/10.1103/PhysRevB.106.L220406

\bibitem{Sorensen} S{\o}rensen, E. S., Catuneanu, A., Gordon, J. S. $\&$ Kee, H.-Y.,
Heart of Entanglement: Chiral, Nematic, and Incommensurate Phases in the Kitaev-Gamma Ladder in a Field.
\emph{Phys. Rev. X} {\bf 11}, 011013 (2021).
https://doi.org/10.1103/PhysRevX.11.011013

\bibitem{Li} Li, H. \etal,
Identification of magnetic interactions and high-field quantum spin liquid in $\alpha$-RuCl$_3$.
\emph{Nat. Commun.} {\bf 12}, 4007 (2021).
https://doi.org/10.1038/s41467-021-24257-8

\bibitem{Chern} Chern, L. E., Kaneko, R., Lee, H.-Y. $\&$ Kim, Y. B.,
Magnetic field induced competing phases in spin-orbital entangled Kitaev magnets.
\emph{Phys. Rev. Research} {\bf 2}, 013014 (2020).
https://doi.org/10.1103/PhysRevResearch.2.013014

\bibitem{Lin2022} Lin, G. \etal,
Evidence for field induced quantum spin liquid behavior in a spin-1/2 honeycomb magnet.
\emph{online preprint at Research Square}
https://doi.org/10.21203/rs.3.rs-2034295/v1.

\end{thebibliography}
\end{document}